\documentclass[JHEP,12pt]{article}
\usepackage{jheppub}
\usepackage{amssymb,amsmath,amsthm}
\usepackage[shortlabels]{enumitem}
\usepackage{braket}
\usepackage{breqn,hyperref}
\usepackage{comment} 
\usepackage{subcaption}
\usepackage{graphicx, xcolor, varwidth}
\usepackage{color}
\newtheorem{thm}{Theorem}
\theoremstyle{definition}
\newtheorem{defn}[thm]{Definition}
 
\newtheorem{lem}[thm]{Lemma}
\theoremstyle{remark}
\newtheorem{rem}[thm]{Remark}

\newtheorem{conv}[thm]{Convention}

\DeclareMathOperator{\setint }{int}

\DeclareMathOperator{\A}{Area}

\newcommand{\emax}{e_{\rm max}}
\newcommand{\emin}{e_{\rm min}}
\newcommand{\smax}{\Sigma_{\rm max}}
\newcommand{\smin}{\Sigma_{\rm min}'}

\title{Geometric Quantum States Beyond AdS/CFT}
\author{Raphael Bousso and Sami Kaya}

\affiliation{Center for Theoretical Physics and Department of Physics,\\
University of California, Berkeley, California 94720, U.S.A. 
} 

\emailAdd{bousso@berkeley.edu}
\emailAdd{samikaya@berkeley.edu}

\abstract{We characterize the quantum states dual to entanglement wedges in arbitrary spacetimes, in settings where the matter entropy can be neglected compared to the geometric entropy. In AdS/CFT, such states obey special entropy inequalities known as the holographic entropy cone. In particular, the mutual information of CFT subregions is monogamous (MMI). We extend this result to arbitrary spacetimes, using a recent proposal for the generalized entanglement wedge $e(a)$ of a gravitating region $a$. Given independent input regions $a$, $b$, and $c$, we prove MMI: $\A[e(a)]+\A[e(b)]+\A[e(c)]-\A[e(ab)]-\A[e(bc)]-\A[e(ca)]+\A[e(abc)]\leq 0$. We expect that the full holographic entropy cone can be extended to arbitrary spacetimes using similar methods.}

\makeatletter
\gdef\@fpheader{\mbox{}}
\makeatother

\begin{document}
\maketitle
\section{Introduction}
Holography has emerged as a powerful framework guiding our search for quantum gravity. Initially, the holographic principle referred to the observation that the matter entropy in a gravitating spatial region is sometimes bounded by its surface area~\cite{tHooft:1993dmi,Susskind:1994vu,Fischler:1998st} and thus highly subextensive. A covariant formulation (Bousso bound)~\cite{Bousso:1999cb,Bousso:1999xy} replaces spatial regions by lightsheets; in this form, the bound appears to hold for all surfaces in our universe. 

Today, holography has become almost synonymous with the AdS/CFT correspondence~\cite{Maldacena:1997re}. The CFT is a complete theory, not just a principle. It supplies a Hamiltonian that can be used, in principle, to construct a unitary S-matrix for the formation and evaporation of a black hole. But it describes quantum gravity only in asymptotically Anti-de Sitter spacetime, and not, for example, in cosmology.

Thus, the holographic principle is a surprising and general property of semiclassical gravity and the gravitational path integral, which applies in arbitrary spacetimes including our universe. Holography in the sense of AdS/CFT is a much stronger, but also much narrower result.


The two notions of holography intersect when we study the entanglement structure of the fundamental quantum states dual to spacetime regions. In AdS/CFT, to leading order in the $G$ or $1/N$ expansion, the entanglement entropy of a spatial CFT subregion $B$ is given by 
\begin{equation}\label{eq:rt}
    S(B) = \frac{\A[RT(B)]}{4G}~,
\end{equation}
where $G$ is Newton's constant and $\A[RT(B)]$ is the area of the minimal~\cite{Ryu:2006bv} (or, more precisely, the minimal stationary~\cite{Hubeny:2007xt}) surface homologous to $B$ in the AdS spacetime. The homology region enclosed between $B$ and $RT(B)$ is called the entanglement wedge of $B$. It constitutes the gravitating region reconstructible from CFT data on $B$~\cite{Wall:2012uf}.

The RT proposal was initially understood to pertain to AdS/CFT specifically. In fact, however, it follows from an application of the gravitational path integral~\cite{Lewkowycz:2013nqa} similar in spirit to the Gibbons-Hawking computation of the thermal black hole partition function~\cite{Gibbons:1976ue}. Thus it is not tied to an asymptotically-AdS setting. In effect, the RT prescription is a geometric shortcut for computing the $n$-th Renyi entropy from the gravitational path integral and analytically continuing to $n=1$, all in one step. 

Therefore, if the fundamental theory (the CFT) was not already known, the RT proposal should properly be viewed as a prediction for the entropy of the states of an \emph{unknown} quantum gravity theory dual to AdS. Even without explicit CFT computations to compare these predictions to, this interpretation of the RT proposal is supported by its several highly nontrivial properties: the quantity it computes obeys strong subadditivity~\cite{Headrick:2007km,Wall:2012uf}, suggesting that it represents a von Neumann entropy; and the entanglement wedge obeys nesting and complementarity~\cite{Wall:2012uf}, suggesting that it represents a reconstructible region.

Based in part on this observation, Bousso and Penington recently proposed a definition of generalized entanglement wedges that applies in arbitrary spacetimes, including our own universe~\cite{Bousso:2022hlz,Bousso:2023sya}. Generalized entanglement wedges, too, can be proven to satisfy strong subadditivity; moreover, they obey appropriate generalizations of nesting and complementarity. These highly nontrivial properties suggest that they represent reconstructible regions, and that their areas represent von Neumann entropies. In AdS, the proposal reduces to the usual entanglement wedges. In any other setting, the quantum gravity theory whose entropy is computed by generalized entanglement wedges is not known. Thus, generalized entanglement wedges can be used to constrain its structure.

In addition to strong subadditivity, which is obeyed by all quantum states, the RT prescription at leading order satisfies an infinite set of inequalities known as the holographic entropy cone \cite{Bao:2015bfa}. The simplest of these inequalities is the monogamy of mutual information (MMI)~\cite{Hayden:2011ag}: for disjoint boundary regions $A,B,C$,
\begin{equation}
    S(AB)+S(BC)+S(AC) \geq S(A)+S(B)+S(C)+S(ABC)~,
\end{equation}
where $AB$ denotes the union.

Unlike strong subadditivity, these additional inequalities are not universal. They must hold only in states for which the original, leading-order RT prescription given in Eq.~\eqref{eq:rt} approximates $S(B)$ well. When matter is present in the bulk, the RT prescription must be modified~\cite{Faulkner:2013ana,Engelhardt:2014gca} so that $S(B)$ receives a contribution from the matter entropy in the entanglement wedge. This contribution can be made arbitrarily large, with negligible backreaction, by adding dilute matter near the boundary, so it can dominate at any value of $G$ or $1/N$. Thus it would not be correct to say that the cone characterizes holographic theories as a whole. 

Rather, the holographic entropy cone characterizes certain states in the holographic theory: those dual to bulk entanglement wedges whose matter entropy can be neglected compared to $A/4G$. In other words, it captures the specific entanglement structure associated with the emergence of gravitating spacetime itself, divorced from the effects of matter fields.

It is therefore of great interest to understand whether the holographic entropy cone, like the RT prescription, extends beyond AdS/CFT. Do the areas of generalized entanglement wedges obey the same infinite set of inequalities as those of RT surfaces, when the matter entropy can be neglected? Here we initiate a study of this question by proving that generalized entanglement wedges obey the simplest of these inequalities: MMI.

Since the required definitions and proofs are somewhat involved, let us warm up by proving MMI for the special ``static'' case analogous to the original RT proposal. Suppose that the spacetime (which is otherwise arbitrary and in particular need not be asymptotically AdS) contains a time-reflection symmetric Cauchy slice $\Sigma$. In this case the required definitions and the proof of MMI simplify considerably. We begin by reviewing the definition of a generalized entanglement wedge, immediately specializing to the ``classical'' case where matter entropy can be neglected.

\begin{defn}
    In the remainder of the introduction, $\Sigma$ will denote a time-reflection symmetric Cauchy slice. We define $\A(a)$ of any subset $a$ of $\Sigma$ as the area of its boundary, $\partial a$.
\end{defn}
\begin{defn}[Static Generalized Entanglement Wedge, Classical Limit~\cite{Bousso:2022hlz}]\label{def:staticgew}
    Let $a\subset \Sigma$ be open (so that $a$ is a partial Cauchy slice, i.e., a spatial region). The static entanglement wedge $E(a)$ is the open subset of $\Sigma$ that contains $a$, has the same conformal boundary as $a$ (if any), and has the smallest boundary area among all such sets.
\end{defn}
\begin{defn}[Static Wedge Union~\cite{Bousso:2022hlz}]
    Let $a,b\subset \Sigma$ be open, with boundaries $\partial a$ and $\partial b$ in $\Sigma$. The wedge union of $a$ and $b$ is 
    \begin{equation}
        a\Cup b\equiv a\cup b\cup (\partial a\cap\partial b)~.
    \end{equation}
    Whenever possible, we use the abbreviated notation
    \begin{equation}
        ab\equiv a\Cup b~.
    \end{equation}
\end{defn}

\begin{thm}[MMI for generalized entanglement wedges, static case] \label{MMIstatic}
Let $a$, $b$ and $c$ be open subsets of $\Sigma$ such that\footnote{In AdS/CFT, the CFT degrees of freedom in disjoint subregions of a boundary Cauchy slice are mutually independent. However, generalized entanglement wedge reconstruction implies that the disjoint gravitating regions $a,b,c$ need not be fundamentally independent~\cite{Bousso:2022hlz,Bousso:2023sya}. Eq.~\eqref{eq:staticindep} provides a suitable notion of independence.}
\begin{equation}\label{eq:staticindep}
    a\cap E(bc) = b\cap E(ca) = c\cap E(ab) =\varnothing~.
\end{equation}
Then
\begin{align}
    \A[E(a)] + \A[E(b)] + \A[E(c)]\nonumber \\ -\A[E(ab)]-\A[E(bc)]- \A[E(ca)]\nonumber \\ + \A[E(abc)] \leq 0~.\label{eq:staticmmi}
\end{align}
\end{thm}

\begin{proof}    
Rearranging components of the areas as in figure \ref{fig1}:
\begin{figure}[h!]
\includegraphics[width=\textwidth]{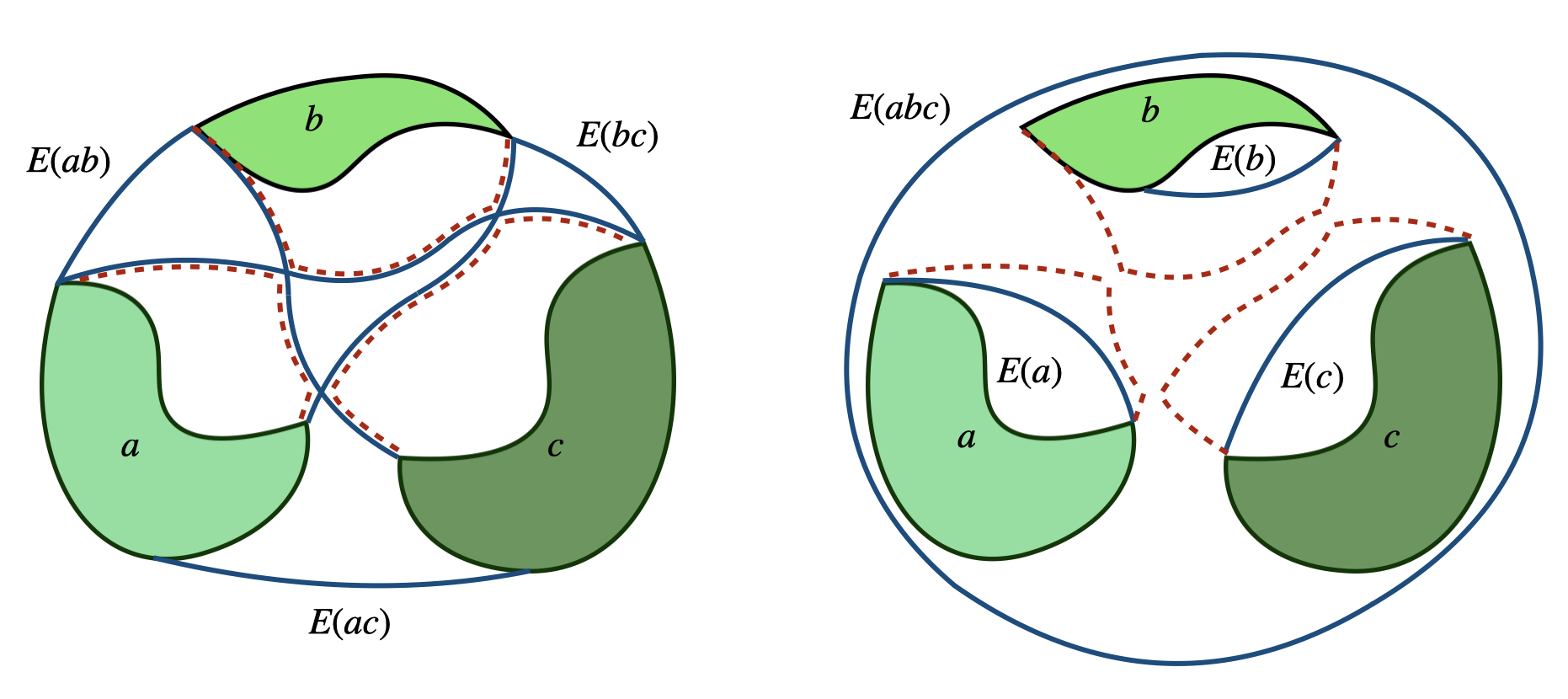}
\caption{\label{fig1} The proof of MMI proceeds in two major steps, shown here for the special case where all relevant regions lie on a time-reflection symmetric Cauchy slice. Left: area portions of $E(ab)$, $E(bc)$, and $E(ca)$ can be rearranged to form the boundaries of 4 new regions, yielding Eq.~(1.7). (The dashed red lines indicate portions of the boundaries of the first 3 regions appearing on the right hand side of that inequality.) Right: defining properties of then  entanglement wedge imply that the areas decrease further when these new regions are deformed into $E(a)$, $E(b)$, $E(c)$, and $E(abc)$.}
\end{figure}
\begin{gather}\label{eq:staticfirst}
\A[E(a b) ]
+ \A[E(a c)]
+ \A[E(b c)] \notag\geq\\  
\A[(E(a b) \cap E(a c)) 
 \setminus E(b c) ]
+ \A[(E(b c) \cap E(b a)) \setminus E(a c) ] \notag \\
+ \A[(E(c b) \cap E(c a)) 
 \setminus E(a b) ] 
+ \A[E(a b) \Cup E(a c) \Cup E(b c))]  ~.
\end{gather}
(This need not be an equality since the wedge union can erase boundary portions.) By Def.~\ref{def:staticgew} (inclusion) and Eq.~\eqref{eq:staticindep}, $a \subset E(ab) \cap E(a c) \setminus E(b c)$. Similarly $b$ and $c$ are contained, respectively, in the second and third set on the right hand side. Again by Def.~\ref{def:staticgew} (inclusion), $abc \subset E(ab) \Cup E(ac) \Cup E(bc)$. By Def.~\ref{def:staticgew} (area-minimization),
\begin{gather}
\A[(E(a b) \cap E(a c)) 
 \setminus E(b c) ]
+ \A[(E(b c) \cap E(b a)) \setminus E(a c) ] \notag \\
+ \A[(E(c b) \cap E(c a)) 
 \setminus E(a b) ] 
+ \A[E(a b) \Cup E(a c) \Cup E(b c))]  \notag
\geq \\\A[E(a)] + \A[E(b)]
+ \A[E(c))]+ \A[E(a b  c)]~.
\end{gather}
\end{proof}

In the remainder of this paper, we will generalize this result to generalized entanglement wedges in time-dependent settings. (This is analogous to extending~\cite{Wall:2012uf} MMI of AdS boundary regions~\cite{Hayden:2016cfa} from the original static context of the RT prescription~\cite{Ryu:2006bv} to the time-dependent setting of the Hubeny-Rangamani-Takayanagi prescription~\cite{Hubeny:2007xt}; we will now do this for generalized entanglement wedges~\cite{Bousso:2022hlz,Bousso:2023sya}.) In Sec.~\ref{definitions}, we reproduce relevant definitions from Ref.~\cite{Bousso:2023sya} and fix notation. In Sec.~\ref{proof} prove MMI for generalized entanglement wedges, building on some novel definitions and Lemmas. 

Together with strong subadditivity~\cite{Bousso:2023sya}, our result establishes the holographic entropy cone for $n\leq 4$ independent subregions. We leave the derivation of the full cone for generalized entanglement wedges to future work.\footnote{Interestingly, aside from MMI, the holographic entropy cone for the time-dependent case (HRT) has not been established even in AdS/CFT. The maximin proof~\cite{Wall:2012uf} of MMI in AdS/CFT does not generalize to more complicated inequalities; see \href{https://www.youtube.com/watch?v=070F7DPoXgE&list=PL9BXLQ4wcldGKcnmzfXpbEh7Fzy-EXUFU&t=872s}{this talk by M.~Headrick}. It will be interesting to study whether the proof techniques we use here are better adapted to this task. Our proof methods differ substantially as they pertain to generalized entanglement wedges where maximin is not useful in any case.}
 
\section{Preliminary Definitions}\label{definitions}

\begin{defn}
  Let $(M,g)$ be a globally hyperbolic manifold with Lorentzian metric, and let $s\subset M$. The domain of influence of $s$, $I(s)$, is the union of $s$ with the set of points that can be reached from some point $p\in s$ by a timelike curve. The domain of dependence of $s$, $D(s)$, is the set of points $q$ such that every causal (i.e., timelike or null) curve through $q$ intersects $s$.
\end{defn}

\begin{defn}\label{def:sc}
The \emph{spacelike complement} of a set $s\subset M$ is defined by
\begin{equation}
    s'\equiv \setint [M\setminus I(s)]~.
\end{equation}
where $\setint$ denotes the interior of a set. (Thus, $s'$ is necessarily open.)
\end{defn}

\begin{defn}\label{def:covwedge}
A {\em wedge} is a set $a\subset M$ that satisfies $a=a''$. 
\end{defn}

\begin{rem}\label{wilem}
Let $a$ be a wedge. By Def.~\ref{def:sc}, $a$ is an open set; $a'$ is also a wedge; and the intersection of two wedges $a,b$ is a wedge.
\end{rem}

\begin{defn}\label{def:edgehor}
The \emph{edge} $\eth a$ and \emph{Cauchy horizon} $H(a)$ of a wedge $a$ are
defined by 
\begin{align}
    \eth a & \equiv \partial a \cap \partial a'~,\\
    H(a) & \equiv \partial D(a)\setminus \eth a~.
\end{align}
where $\partial$ denotes the boundary of a set. It will also be convenient to define
\begin{align}
    \overline{a} & \equiv a \sqcup H(a)~;
\end{align}
we use the symbol $\sqcup$ to indicate that the union is disjoint.
\end{defn}

\begin{rem}
Any wedge $a$ induces a decomposition of the spacetime $M$ into disjoint sets:
\begin{align}\label{eq:decom}
    M & = a \sqcup a' \sqcup I(\eth a) \sqcup H(a) \sqcup H(a') \\
    & = \overline a \sqcup \overline{a'} \sqcup I(\eth a)~.
\end{align}
\end{rem}

\begin{defn}
The {\em wedge union} of two wedges $a,b$ is the wedge
\begin{equation}
    a\Cup b\equiv (a'\cap b')'~.
\end{equation}
As in the static case, we will suppress the symbol $\Cup$ and simply write $ab$ whenever possible.
\end{defn}

\begin{defn}
The area of a wedge $a$ is defined as the area of its edge. 
\end{defn}

\begin{rem}
    By Def.~\ref{def:edgehor}, $\A(a)=\A(a')$.
\end{rem}

\begin{figure}[h!]
\centering
\includegraphics[width=.6\textwidth]{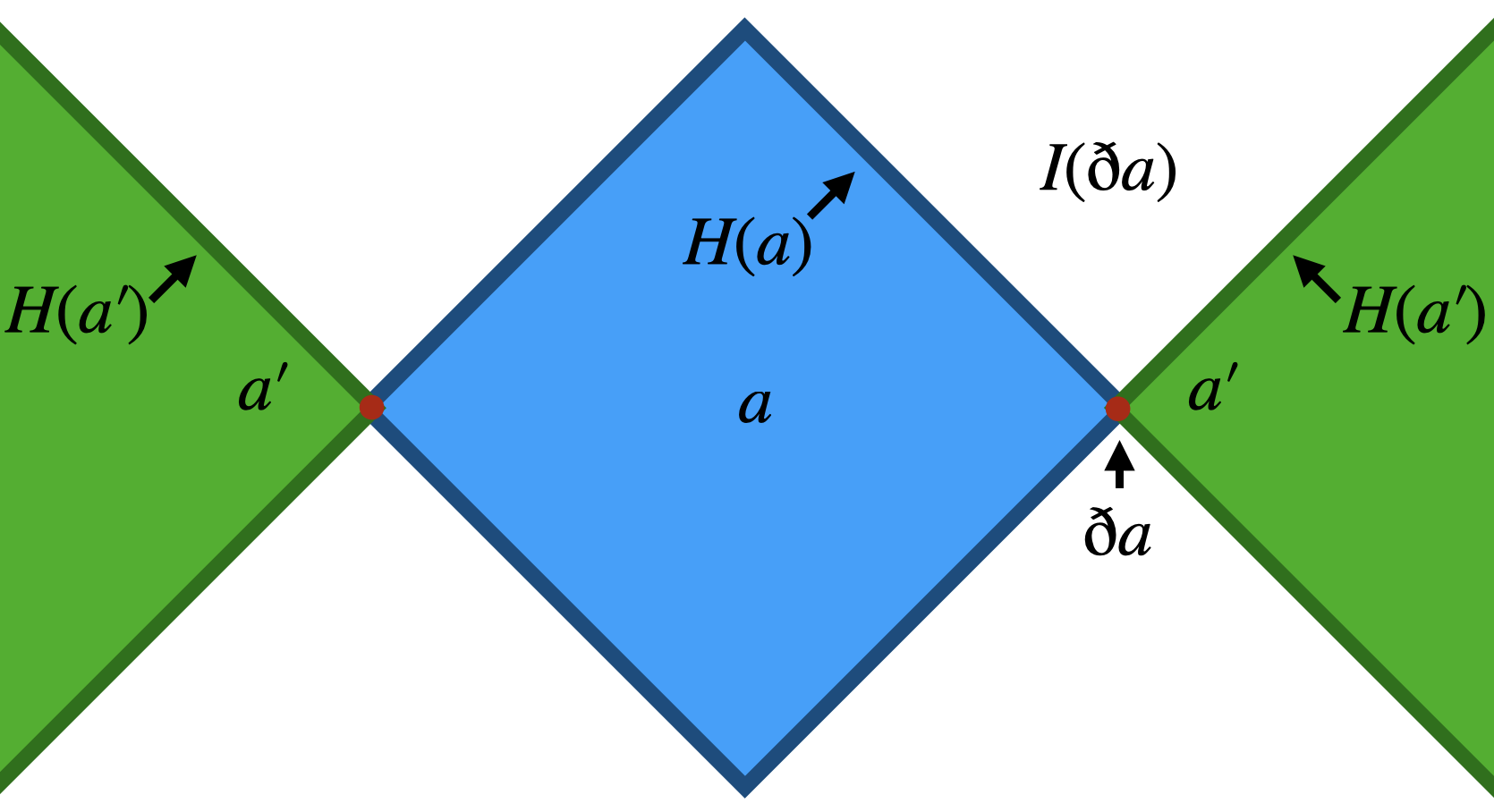}
\caption{\label{adecomp} A wedge $a$ induces a decomposition of the spacetime as shown. Here $H(a)$ is the Cauchy horizon of $a$. $I(\eth a)$ is the domain of influence of the edge $\eth a$ of $a$. The remaining part of the spacetime consists of the complement wedge $a'$ and its Cauchy horizon $H(a')$.}
\end{figure}


\begin{defn}
Let $\theta^\pm(a,p)$ be the expansions~\cite{Wald:1984rg} at $p$ of the null congruences orthogonal to $\eth a$ that enter $H^\pm(a')$. The wedge $a$ is called \emph{extremal}\, at $p\in \eth a$ if $\theta^+(a,p)=\theta^-(a,p)=0$. Similarly, the wedge is called \emph{normal} at $p$ if both expansions are positive or zero, and \emph{antinormal} if both are negative or zero.
\end{defn}

\begin{defn}
The spacetime $(M,g)$ is \emph{weakly classically focussing} (WCF) if the expansion of a null congruence is nonincreasing at all points where it vanishes. 
\end{defn}

\begin{rem}
The WCF property is a classical version of the restricted~\cite{Shahbazi-Moghaddam:2022hbw} Quantum Focussing Conjecture~\cite{Bousso:2015mna}. It holds in particular if the null curvature condition is satisfied on $(M,g)$, and hence if Einstein's equations are satisfied with the null energy condition holding for matter.
\end{rem}

\begin{defn}[$\emax$]\label{emaxdef}
  Given a wedge $a$, let $F(a)\equiv \set{f:\mathrm{I}\,\wedge\,\mathrm{II}\,\wedge\,\mathrm{III}}$ be the set of all wedges that satisfy the following properties:
  \begin{enumerate}[I.]
  \item $f\supset a$ and $\tilde \eth f=\tilde \eth a$~;
  \item $f$ is antinormal at points $p\in \eth f\setminus\eth a$~;
  \item $f$ admits a Cauchy slice $\Sigma$ such that
    \begin{enumerate}
    \item $\Sigma\supset \eth a$~;
    \item $\A(h) > \A(f)$ for any wedge $h \neq f$ such that $a\subset h$, $\eth h\subset \Sigma$, and $\eth h\setminus \eth f$ is compact in $M$.
    \end{enumerate}
   \end{enumerate}
  The {\em classical max-entanglement wedge} of $a$, $\emax(a)$, is their wedge union:
   \begin{equation}\label{eq:emaxdef}
       \emax(a) \equiv \Cup_{f\in F(a)}\, f~.
   \end{equation}
\end{defn}

\begin{defn}[$\emin$]\label{emindef}
  Given a wedge $a$, let $G(a)\equiv \set{g: \mathrm{i}\, \wedge\, \mathrm{ii} \,\wedge\, \mathrm{iii}}$ be the set of all wedges that satisfy the following properties:
  \begin{enumerate}[i.]
  \item $g\supset a$~;\footnote{In Ref.~\cite{Bousso:2023sya}, the additional condition $\tilde \eth f=\tilde \eth a$ was required as part of property i. However, this condition interferes with the proof that $\emin(a)\in G(a)$ (Theorem 23 of \cite{Bousso:2023sya}; specifically one needs that $M\in G(a)$). In fact, under reasonable conditions on the asymptotic structure, the condition $\tilde \eth f=\tilde \eth a$ is unnecessary, since wedges in $G(a)$ with a larger conformal boundary will contain subwedges with $\tilde \eth f=\tilde \eth a$ that satisfy properties i--iii.}. 
  \item $g$ is normal; 
  \item $g'$ admits a Cauchy slice $\Sigma'$ such that $\A(h) > \A(g)$ for any wedge $h \neq g$ such that $g \subset h$, $\eth h\subset \Sigma'$, and $\eth h\setminus \eth g$ is compact.
  \end{enumerate}
  The {\em classical min-entanglement wedge} of $a$, $\emin(a)$, is their intersection:
  \begin{equation}\label{eq:emindef}
      \emin(a)\equiv \cap_{g\in G(a)}\, g~.
  \end{equation}
\end{defn}

\begin{rem}
    In general, $\emax(a)$ and $\emin(a)$ need not agree. (This is also true in the special case of entanglement wedges of conformal boundary regions in AdS/CFT. However, in that case they can only disagree if the bulk matter is in an incompressible quantum state~\cite{Akers:2020pmf,Akers:2023fqr}, whereas more generally they can already disagree at the classical level.)  If $\emax(a)=\emin(a)$, we will denote them as $e(a)$. In this case, $\A[e(a)]$ satisfies strong subadditivity and is expected to represent a von Neumann entropy in a fundamental quantum gravity theory~\cite{Bousso:2023sya}.
\end{rem}

\section{Proof of MMI}\label{proof}

\begin{defn}
Let $a$ be a wedge, and let $p\in \eth a$. If $\eth a$ is a $C^1$ submanifold at $p$, there are 4 null geodesics that begin at $p$ and are orthogonal to $\eth a$. We define $\gamma_p^+(a)$, $\gamma_p^-(a)$, $\gamma_p^+(a')$, and $\gamma_p^-(a')$ to be their intersection with $H^+(a)$, $H^-(a)$, $H^+(a')$, and $H^-(a')$, respectively; see Fig.~FFF. We further define the broken null geodesics
\begin{align}
    \gamma_p(a) & \equiv \gamma_p^+(a) \sqcup {p} \sqcup \gamma_p^-(a)~;\\
    \gamma_p(a') & \equiv \gamma_p^+(a') \sqcup {p} \sqcup \gamma_p^-(a')~.
\end{align}
\end{defn}
\begin{figure}[h!]
\includegraphics[width=7cm]{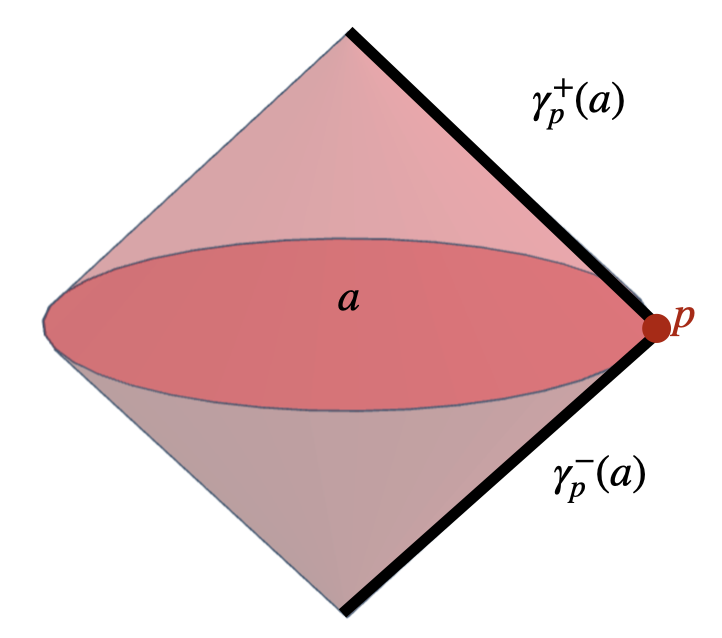}
\centering
\caption{\label{gammadecomp} Every point $p$ on the edge of a wedge is the starting point of two null geodesics $\gamma_p(a)^\pm$ that lie on the past and future Cauchy horizon, $H^\pm(a)$.}
\end{figure}
\begin{defn}
    Let $a$, $b$, and $c$ be wedges. We define
    \begin{equation}\label{eq:abcdef}
        \eth a_{bc}\equiv \set{p\in \eth a\setminus \overline{b'} \setminus \overline{c'} \setminus (\overline{b}\cap\overline{c}) : \gamma_p(a)\cap \overline{b}\cap\overline{c} \neq\varnothing}~.
    \end{equation}
    and
    \begin{equation}\label{eq:abcpdef}
        \eth a'_{bc}\equiv \set{p\in \eth a\setminus \overline{b'} \setminus \overline{c'} \setminus (\overline{b}\cap\overline{c}) : \gamma_p(a')\cap \overline{b}\cap\overline{c} \neq\varnothing}~.
    \end{equation}
\end{defn}

\begin{rem}
    Intuitively, it is helpful to think of $\eth a_{bc}$ as the portion of the edge of $a$ whose orthogonal null geodesics towards $a$ enter $b\cap c$. For example, generically Eq.~\eqref{eq:abcdef} could be replaced by $\eth a_{bc}\equiv \set{p\in \eth a\setminus (b\cap c) : \gamma_p(a)\cap b\cap c \neq\varnothing}$. The above definition is designed to handle marginal cases correctly, for example when $\eth a\subset H[(b\cap c)']$. Accounting for these cases is tedious but straightforward, and we will not spell out all the possible marginal configurations in proofs. On a first reading, we recommend making a generic assumption about the spacetime, so that such cases will not occur.
\end{rem}

\begin{lem}\label{lem:abcdisjoint}
    Let $a$, $b$, and $c$ be wedges. Then the four sets $\eth a_{bc}$, $\eth a_{bc'}$, $\eth a_{b'c}$, and $\eth a_{b'c'}$ are mutually disjoint. Moreover, the four sets $\eth a'_{bc}$, $\eth a'_{bc'}$, $\eth a'_{b'c}$, and $\eth a'_{b'c'}$ are mutually disjoint.
\end{lem}

\begin{proof}
    For contradiction, suppose that $p\in \eth a_{bc} \cap \eth a_{bc'}$. By the first portion of Eq.~\eqref{eq:abcdef}, 
    \begin{equation}\label{eq:pnotinstuff}
        p\notin \overline{c}\cup\overline{c'}~.
    \end{equation} 
    We now recall Def.~\ref{def:edgehor}, that $\overline{c}=c\cup H(c)$ and $\overline{c'}=c'\cup H(c')$. By the second portion of Eq.~\eqref{eq:abcdef} (the criterion after the colon), $\gamma_p(a)$ intersects $\overline{c}$ and also $\overline{c'}$. By Eq.~\eqref{eq:decom} and Theorem 9.3.11 of Ref.~\cite{Wald:1984rg} (see also Ref.~\cite{Akers:2017nrr}), this requires that $\gamma_p(a)$ contain an unbroken null geodesic with endpoints $p_1\in \overline{c}$ and $p_2 \in \overline{c'}$. Since $\gamma_p$ is broken at $p$, this implies $p\in \overline{c} \cup \overline{c'}$; this contradicts Eq.~\eqref{eq:pnotinstuff}. Similarly one shows that that all other pairs among the first four sets are disjoint, and that all of the second four sets are mutually disjoint.
    
\end{proof}

\begin{lem} \label{3intersect}
    Let $a$, $b$, and $c$ be wedges, and suppose that they are normal at points on $\eth a_{bc}$, $\eth b_{ca}$, and $\eth c_{ab}$. Then
    \begin{align}\label{eq:3intersect}
        \A(a\cap b\cap c) \leq & \A(\eth a \cap \overline{b} \cap \overline{c}) + \A(\eth b \cap \overline{c} \cap \overline{a})+ \A(\eth c \cap \overline{a} \cap \overline{b}) \nonumber\\ & +\frac{1}{2} \left[ \A(\eth a_{bc}) +\A(\eth b_{ca}) +\A(\eth c_{ab})\right]~.
    \end{align}
\end{lem}

\begin{proof}
    Every point in $\eth(a\cap b\cap c)$ lies either in one of the sets appearing in the first line, or on the intersection of two broken null geodesics $\gamma$ originating from two distinct sets of the second line.     
    (See Fig.~\ref{fig:intersectgammas} for an example.) 
    \begin{figure}[h!]
\includegraphics[width=4.6 cm]{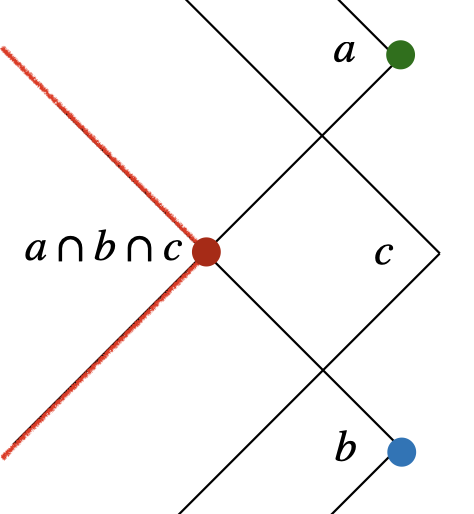}
\centering
\caption{\label{fig:intersectgammas} Example illustrating the second line of Eq.~\eqref{eq:3intersect}. Here, the set $\set{p\in \eth(a\cap b\cap c): p=\gamma_q(a)\cap\gamma_r(b)~,~q\in \eth a_{bc}~,~r \in \eth b_{ca}}$ (red dot) forms a portion of $\eth(a\cap b\cap c)$. By the weak focusing condition, the area of this portion is smaller than that of either ``origin'' set, $\eth a_{bc}$ (green dot) and $\eth b_{ca}$ (blue dot), and hence smaller than or equal to half their sum.}
\end{figure}
\end{proof}

\begin{lem} \label{lem:3areas}
    Let $a$ and $b$, and $c$ be wedges, and suppose that they are normal at points on 
    $\eth b_{ca'} $, $\eth c_{a'b}
    $, $\eth a_{b'c} $, $\eth c_{ab'}
    $, $\eth a_{bc'} $, and $\eth b_{c'a} $; and antinormal on $\eth a'_{bc} $, $\eth b'_{ca} $, $\eth c'_{ab}
    $, $\eth a'_{b'c'} $, $\eth b'_{c'a'} $, and $\eth c'_{a'b'}$. Then
    \begin{align}
        \A(a'\cap b\cap c) + \A(a\cap b'\cap c) +\A(a\cap b\cap c') +\A(a'\cap b'\cap c') \nonumber \\ \leq \A(a)+\A(b)+\A(c)~.
    \end{align}
\end{lem}
\begin{proof}
By the preceding Lemma,
\begin{align}
    \A(a'\cap b\cap c) + \A(a\cap b'\cap c) +\A(a\cap b\cap c') +\A(a'\cap b'\cap c') \nonumber \\
    \leq \A(\eth a \cap \overline{b} \cap \overline{c}) 
    + \A(\eth b \cap \overline{c} \cap \overline{a'})+ \A(\eth c \cap \overline{a'} \cap \overline{b}) \nonumber\\
    +\A(\eth a \cap \overline{b'} \cap \overline{c}) 
    + \A(\eth b \cap \overline{c} \cap \overline{a})+ \A(\eth c \cap \overline{a} \cap \overline{b'}) \nonumber\\
    +\A(\eth a \cap \overline{b} \cap \overline{c'}) 
    + \A(\eth b \cap \overline{c'} \cap \overline{a})+ \A(\eth c \cap \overline{a} \cap \overline{b}) \nonumber\\
    +\A(\eth a \cap \overline{b'} \cap \overline{c'}) 
    + \A(\eth b \cap \overline{c'} \cap \overline{a'})+ \A(\eth c \cap \overline{a'} \cap \overline{b'}) \nonumber\\
    +\frac{1}{2}\left[ \A(\eth a'_{bc}) +\A(\eth b_{ca'}) +\A(\eth c_{a'b})\right. \nonumber\\
    +\A(\eth a_{b'c}) +\A(\eth b'_{ca}) +\A(\eth c_{ab'})\nonumber\\
    +\A(\eth a_{bc'}) +\A(\eth b_{c'a}) +\A(\eth c'_{ab})\nonumber\\
    \left.+\A(\eth a'_{b'c'}) +\A(\eth b'_{c'a'}) +\A(\eth c'_{a'b'}) \right]
\end{align}
Consider the 8 portions of $\eth a$ that appear on the right hand side; they correspond to the first term on each of the 8 lines. The first 4 of these terms are mutually disjoint; and their union is disjoint from all of the remaining 4 terms. By Lemma~\ref{lem:abcdisjoint} these remaining 4 terms are pairwise disjoint: 
\begin{equation}
    \eth a_{b'c}\cap \eth a_{bc'} = \varnothing~;~~~ 
    \eth a'_{bc}\cap \eth a'_{b'c'} = \varnothing~.
\end{equation}
It follows that any point $p\in \eth a$ that lies outside the first four portions appears \emph{at most twice} in the remaining four portions. The factor of $1/2$ ensures that the sum of the areas of the first term in each line will not exceed $\A(a)$: 
\begin{align}
    \A(\eth a \cap \overline{b} \cap \overline{c}) 
    +\A(\eth a \cap \overline{b'} \cap \overline{c}) 
    +\A(\eth a \cap \overline{b} \cap \overline{c'}) 
    +\A(\eth a \cap \overline{b'} \cap \overline{c'}) \nonumber \\
    +\frac{1}{2}\left[ \A(\eth a'_{bc}) +\A(\eth a_{b'c})+\A(\eth a_{bc'})
    +\A(\eth a'_{b'c'})\right]\leq \A(a)~.
\end{align}
Similarly one finds that the second and third terms of each line contribute less area than $\A(b)$ and $\A(c)$, respectively. 
\end{proof}

\begin{conv}\label{conv:xyz}
    In the remainder of this paper, let $x$, $y$, and $z$ be wedges, such that $\emin=\emax$ for $x$, $y$, $z$ and for all their wedge unions. We can thus denote the entanglement wedges by $e$ and define $a=e(yz)$, $b=e(zx)$, and $c=e(xy)$. We will assume that they satisfy the independence conditions $x\subset a'$, $y\subset b'$, and $z\subset c'$.
\end{conv}

\begin{lem}\label{lem:sat}
    $a$, $b$, and $c$ as defined in Convention~\ref{conv:xyz} satisfy the assumptions of Lemma~\ref{lem:3areas}.
\end{lem}

\begin{proof}
    Since $a$, $b$ and $c$ are min-entanglement wedges, they are everywhere normal. It remains to be shown that they are extremal (and hence anti-normal) on $\eth a'_{bc} $, $\eth b'_{ca} $, $\eth c'_{ab}
    $, $\eth a'_{b'c'} $, $\eth b'_{c'a'} $, and $\eth c'_{a'b'}$. We will show this for $\eth a'_{bc}$ and $\eth a'_{b'c'}$; the proofs for the remaining 4 sets obtain by cyclic permutation.

    By Eq.~\eqref{eq:abcpdef}, $\eth a'_{bc}\cap \overline{b'}=\varnothing$ and $\eth a'_{bc}\cap \overline{c'}=\varnothing$. The independence assumption in Convention~\ref{conv:xyz} implies $y\subset b'$ and $z\subset c'$. Hence $\eth a'_{bc}\cap \eth y=\varnothing$ and $\eth a'_{bc}\cap \eth z=\varnothing$. While $y\subset b'$, Def.~\ref{emaxdef} implies that $z\subset b$; hence $y$ and $z$ are mutually spacelike. It follows that $\eth(yz)\subset \eth y \cup \eth z$. Combining these results we find that $\eth a'_{bc}\cap \eth(yz)=\varnothing$. By Lemma 4.14 of Ref.~\cite{Bousso:2022hlz}, $\eth a'_{bc}$ is extremal.

    Similarly, $\eth a'_{b'c'}\cap \overline{b}=\varnothing$, $\eth a'_{b'c'}\cap \overline{c}=\varnothing$, $y\subset c$, and $z\subset b$ imply that $\eth a'_{b'c'}$ is extremal.
\end{proof}

\begin{lem}\label{lem:normal}
    With $a$, $b$, and $c$ defined as in Convention~\ref{conv:xyz}, the wedges $a'\cap b \cap c$, $a\cap b' \cap c$, $a\cap b \cap c'$, and $a'\cap b' \cap c'$ are normal. 
\end{lem}

\begin{proof}
    Suppose for contradiction that $a'\cap b\cap c$ is not normal at the point $q\in \eth(a'\cap b\cap c)$. By virtue of being min-entanglement wedges, $b$, and $c$ are normal. Hence $b\cap c$ is normal. This implies that $q\in b\cap c\cap \gamma_p(a')$. By Theorem 21 of~\cite{Bousso:2023sya}, $a$ is extremal except where its edge overlaps with $\eth y$ or $\eth z$, so weak focusing implies $p\in \eth y\cup \eth z$. (Here we used the fact that $y$ and $z$ are mutually spacelike; see the proof of Lemma~\ref{lem:sat} above.) Hence there exists a causal curve from $b\cap c$ to $y\cup z$. This contradicts at least one of the independence conditions $y\subset b'$, $z\subset c'$. Hence $a'\cap b\cap c$ is normal. By cyclic permutation, $a\cap b'\cap c$ and $a\cap b\cap c'$ are also normal.

    Suppose for contradiction that $a'\cap b'\cap c'$ is not normal at the point $q\in \eth(a'\cap b'\cap c')$. Again using Theorem 21 of~\cite{Bousso:2023sya}, this implies $[I(ab)\cap xy] \cup [I(bc)\cap yz] \cup[I(ca)\cap zx] \neq \varnothing$. But in fact this set must be empty, by the independence conditions of Convention~\ref{conv:xyz} and property i in Def.~\ref{emindef}; the latter applies since  $\emin(x)\in G(x)$) by Theorem 23 of~\cite{Bousso:2023sya}.
\end{proof}

\begin{thm}(Monogamy of Mutual Information): 

With $x$, $y$, and $z$ as in Convention~\ref{conv:xyz},
\begin{gather}
    \A[e(xy)]+ \A[e(yz)]+ \A[e(zx)]  \notag \\ \geq \A[e(x)] + \A[e(y)] +\A[e(z)]+ \A[e(xyz)] 
    \label{eqnmmi}
\end{gather}
\end{thm}
\begin{proof}
By Lemma~\ref{lem:sat} and Lemma~\ref{lem:3areas}, 
\begin{gather}
    \A[e(xy)]+ \A[e(yz)]+ \A[e(zx)]  \notag \\ \geq \A[e(xy)\cap e(yz) \cap e(zx)'] +
    \A[e(xy)\cap e(yz)' \cap e(zx)] \notag\\+
    \A[e(xy)'\cap e(yz) \cap e(zx)]+
    \A[e(xy) \Cup e(yz) \Cup e(zx)] 
    \label{mmi3areas}
\end{gather}
Nesting of $\emin$ (Theorem 27 of~\cite{Bousso:2023sya}) implies $e(x)\subset e(xy)$ and $e(x)\subset e(zx)$. The independence conditions state that $x\subset e(yz)'$ and imply (using nesting) that $yz\subset e(x)'$; the No-Cloning (Theorem 29 of~\cite{Bousso:2023sya}) then implies $e(x)\subset e(yz)'$. Combining these results, we find that
\begin{align}
    e(x) &\subset e(xy)\cap e(yz)'\cap e(zx)~,\\
    e(y) &\subset e(yz)\cap e(zx)'\cap e(xy)~,\\
    e(z) &\subset e(xy)\cap e(yz)'\cap e(zx)~,
\end{align}
where the second and third line follow by cyclic permutation. By Theorem 23 of~\cite{Bousso:2023sya}, $\emin(x)\in G(x)$; let $\smin(x)$ be the Cauchy surface of $\emin(x)'$ guaranteed to exist by Property iii of Def.~\ref{emindef}. Property iii and Lemma~\ref{lem:normal} imply
\begin{align}
    \A[e(x)] \leq \A[e(xy)\cap e(yz)'\cap e(zx)\cap \smin(x)] 
    \leq  \A[e(xy)\cap e(yz)'\cap e(zx)]~;
\end{align}
and cyclic permutation yields analogous inequalities for $\A[e(y)]$ and $\A[e(z)]$.

Nesting also implies
\begin{equation}
    e(xyz) \supset e(xy)\Cup e(yz)\Cup e(zx)~.
\end{equation}
By Theorem 19 of~\cite{Bousso:2023sya}, $\emax(xyz)\in F(xyz)$; let $\smax(xyz)$ be the Cauchy surface of $\emax(xyz)$ guaranteed to exist by Property III of Def.~\ref{emaxdef}. Property III and Lemma~\ref{lem:normal} imply
\begin{align}
    \A[e(xyz)] \leq \A[e(xy)'\cap e(yz)'\cap e(zx)'\cap \smax(x)] 
    \leq  \A[e(xy)\Cup e(yz)\Cup e(zx)]~.
\end{align}
Combining all of the above area inequalities yields Eq.~\eqref{eqnmmi}. 
\end{proof}

\subsection*{Acknowledgements}
We thank Chris Akers, Brianna Grado-White, and Pratik Rath for discussions. This work was supported in part by the Berkeley Center for Theoretical Physics; and by the Department of Energy, Office of Science, Office of High Energy Physics under QuantISED Award DE-SC0019380.

\bibliographystyle{JHEP}
\bibliography{covariant1}
\end{document}